\documentclass[conference,twocolumn]{IEEEtran}
\usepackage[final]{graphicx}
\DeclareGraphicsExtensions{.eps}

\usepackage[cmex10]{amsmath}
\usepackage{fancyhdr}
\usepackage{amsmath,epsfig,bm,amssymb,amsthm}
\usepackage{psfrag,accents}
\usepackage{cite}
\usepackage{color}
\ifCLASSINFOpdf
\else
\fi

\usepackage{amsmath,epsfig,bm,amssymb,amsthm,balance}

\newcommand{\sinc}{{\mathrm {sinc}}}
\newcommand{\et}{{\mathrm {e}}}
\newcommand{\rt}{{\mathrm {r}}}
\newcommand{\SR}{{\mathrm {SR}}}
\newcommand{\RiD}{{\mathrm {R}_i\mathrm{D}}}

\newcommand{\IDFT}{{\mathrm {IDFT}}}
\newcommand{\DFT}{{\mathrm {DFT}}}

\newcommand{\Zc}{\mathcal{Z}}

\newcommand{\Vb}{\mathbf{V}}

\newcommand{\s}{\mathbf{s}}

\newcommand{\0}{\mathbf{0}}

\newcommand{\y}{\mathbf{y}}

\newcommand{\w}{\mathbf{w}}
\newcommand{\CN}{\mathcal{CN}}
\newcommand{\Vc}{\mathcal{V}}

\newcommand{\I}{\mathbf{I}}

\newcommand{\addt}[1]{\textcolor{black}{{#1}}}

\IEEEoverridecommandlockouts

\begin{document}


\title{Differential Distributed Space-Time Coding \\ with Imperfect Synchronization}

\author{
	\IEEEauthorblockN{M. R. Avendi, Sina Poorkasmaei and Hamid Jafarkhani \\
Center for Pervasive Communications and Computing, University of California, Irvine, USA
\\
Email: \{m.avendi, spoorkas, hamidj\}@uci.edu
\thanks{This work was supported in part by the NSF Award CCF-1218771.}
}
}

\maketitle

\begin{abstract}
\label{abs}
Differential distributed space-time coding (D-DSTC) has been considered to improve both diversity and data-rate in cooperative communications in the absence of channel information. However, conventionally, it is assumed that relays are perfectly synchronized in the symbol level. In practice, this assumption is easily violated due to the distributed nature of the relay networks. This paper proposes a new differential encoding and decoding process for D-DSTC systems with two relays. The proposed method is robust against synchronization errors and does not require any channel information at the destination. Moreover, the maximum possible diversity and symbol-by-symbol decoding are attained. Simulation results are provided to show the performance of the proposed method for various synchronization errors and the fact that our algorithm is not sensitive to synchronization error.
\end{abstract}

\begin{keywords}
Distributed space-time coding, differential encoding and decoding, synchronization error, OFDM, relay networks
\end{keywords}

\IEEEpeerreviewmaketitle
\lhead{IEEE Globecom, 2014}

\section{Introduction}
\label{sec:intro}
Cooperative communication techniques make use of the fact that, since users in a network can listen to a source during its transmission phase, they would be able to re-broadcast the received data to the destination in another phase. Therefore, the overall diversity and performance of a network would benefit from a virtual antenna array that is constructed cooperatively by multiple users.

Depending on the protocol that relays utilize to process and re-transmit the received signal to the destination, relay networks have been generally classified as decode-and-forward or amplify-and-forward \cite{coop-laneman}. Among these two protocols, amplify-and-forward (AF) has been the focus of many studies because of its simple relay operation. Moreover, depending on the strategy that relays utilize to cooperate, relay networks are categorized as repetition-based and distributed space-time coding (DSTC)-based \cite{DSTC-Laneman}. At the price of higher complexity, the latter strategy yields a higher spectral efficiency than the former \cite{DSTC-Laneman}.

In DSTC networks \cite{DSTC-HJ,DSTC-Laneman,DSTC-Y,DSTC-Kaveh}, the relays cooperate to combine the received symbols by multiplying them with a fixed or variable factor and forward the resulting signals to the destination. The cooperation is such that a space-time code is effectively constructed at the destination. Coherent detection of transmitted symbols can be achieved by providing the instantaneous channel state information (CSI) of all transmission links at the destination. Although this requirement can be accomplished by sending pilot (training) signals and using channel estimation techniques, it is not feasible or efficient in relay channels as there are more channels involved in the communication. Moreover, the computational complexity and overhead of channel estimation increase proportionally with the number of relays. Also, all channel estimation techniques are subject to impairments that would directly translate to performance degradation.

When no CSI is available at the relays and destination, differential DSTC (D-DSTC) has been studied in \cite{D-DSTC-Y,D-DSTC-Amin,D-DSTC-Giannakis}. D-DSTC only needs the second-order statistics of the channels at the relays. Also, the constructed unitary space-time code at the destination together with differential encoding provides the opportunity to apply non-coherent detection without any CSI.

On the other hand, due to the distributed nature of relay networks, the received signals from relays at the destination are not always aligned in the symbol level. This, so-called synchronization error between relays, causes inter symbol interference (ISI). For coherent DSTC, synchronization error has been studied in  \cite{ADSTC-Valenti,ADSTC-Hua,ADSTC-HJ,ADSTC-Olafsson,ADSTC-FF-2R-Xia1}. However, their methods require not only the CSI of all channels but also the amount of synchronization error between relays. Also, the previous studies on D-DSTC \cite{D-DSTC-Giannakis,D-DSTC-Amin,D-DSTC-Y} all assume that perfect synchronization in the symbol level exists between relays.  

Based on the above motivations, in this article, a differential encoding and decoding process is designed to combat synchronization error when neither CSI nor synchronization delay are available at the destination. We consider the case that a source communicates with a destination via two relays and the received signals from the two relays may not be aligned. All channels are assumed to be Rayleigh flat-fading and slowly changing over time. The effect of synchronization error is modeled as the effect of frequency-selective channels and differential encoding and decoding are combined with an OFDM approach to circumvent both channel estimation and the ISI. At the source, differential encoding and the Inverse Discrete Fourier Transform (IDFT) are employed. At the relays, essential configuration and a protecting guard are applied as will be detailed later. At the destination, the Discrete Fourier Transform (DFT) and differential decoding are utilized to obtain a symbol-by-symbol decoding with low complexity. The proposed method does not require any CSI or the amount of synchronization error and provides significant performance improvement compared to cases with symbol misalignment. Simulation results show the effectiveness of this method for various values of errors. A performance difference of around 3 dB is seen between our method and that of coherent DSTC with perfect synchronization.

The outline of the paper is as follows. Section \ref{sec:cdd} describes the conventional system model and the problem statement. In Section \ref{sec:DSTC-OFDM}, the proposed method is described. Simulation results are given in Section \ref{sec:Sim}. Section \ref{sec:con} concludes the paper.

\emph{Notations:} $(\cdot)^t$, $(\cdot)^*$ and $|\cdot|$ denote transpose, complex conjugate and absolute value of a complex number, respectively. $\I_R$ and $\0_R$ are the $R \times R$ identity and zero matrices, respectively. $\CN(\0,\sigma^2 \I_R)$ stands for circularly symmetric Gaussian random vector with zero mean and covariance $\sigma^2 \I_R$. $\mbox{E}\{\cdot\}$ denotes the expectation operation. $\| \cdot \|$ denotes the Euclidean norm of a vector. $\otimes$ is the circular convolution operation. $\Zc$ is the set of integer numbers. $\sinc(x)=\sin(\pi x)/(\pi x)$.

\begin{figure}[t]
\psfrag {Source} [] [] [1.0] {Source}
\psfrag {Relay1} [] [] [1.0] {Relay 1}
\psfrag {Relay2} [] [] [1.0] {Relay 2}
\psfrag {Destination} [] [] [1.0] {Destination}
\psfrag {s1} [r] [] [1.0] {}
\psfrag {x1} [r] [] [1.0] {}
\psfrag {x2} [r] [] [1.0] {}
\psfrag {y1} [r] [] [1.0] {}
\psfrag {f1} [r] [] [1.0] {$q_1$}
\psfrag {f2} [bl] [] [1.0] {$q_2$}
\psfrag {g1} [l] [] [1.0] {$g_1$}
\psfrag {g2} [l] [] [1.0] {$g_2$}
\centerline{\epsfig{figure={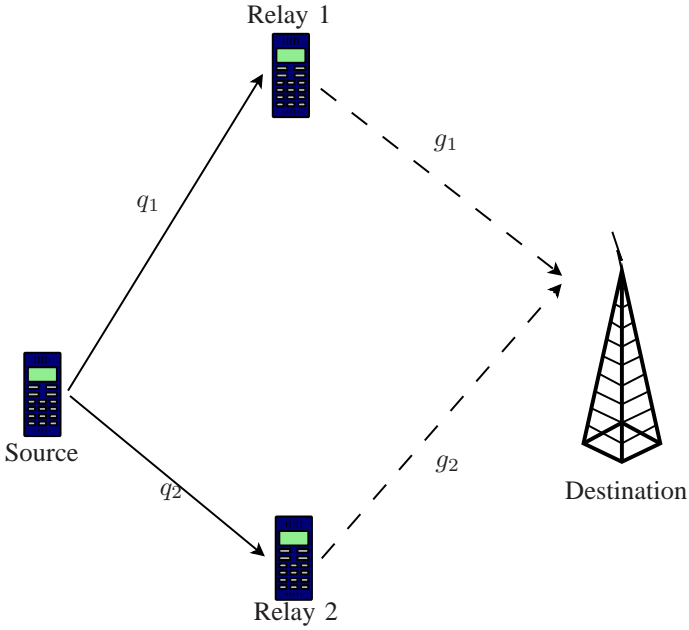},width=8.5cm}}
\caption{Cooperative network under consideration, Source communicates with Destination through two relays.}
\label{fig:sysmodel}
\end{figure}

\section{Conventional System and Problem Statement}
\label{sec:cdd}
In this section the system model based on the conventional D-DSTC system \cite{D-DSTC-Y} is recalled. The wireless relay network under consideration is depicted in Fig.~\ref{fig:sysmodel}. There is one source, two relays and one destination. All nodes have one antenna and transmission is half-duplex (i.e., each node can only transmit or receive at one time). The wireless channels between the nodes are all flat-fading channels. In the conventional D-DSTC system, information bits are converted to symbols from constellation set $\Vc$ (such as PSK, QAM) at Source. Let us assume that two symbols $v_1, v_2 \in \Vc$ are going to be sent from Source to Destination. The transmission process is divided into two phases and sending two symbols from Source to Destination in two phases is referred to as ``one transmission block", indexed by $k\in \mathcal{Z}$. 
First, symbols are encoded to a unitary space-time coding (USTC) matrix as \cite{D-DSTC-Y}
\begin{equation}
\label{eq:alamouti}
\Vb^{(k)}= \frac{1}{\sqrt{|v_1|^2+|v_2|^2}}
\begin{bmatrix}
v_1 & -v_2^* \\
v_2 & v_1^*
\end{bmatrix}. 
\end{equation}
Before transmission, the codeword is differentially encoded as
\begin{equation}
\label{eq:s[k]}
\s^{(k)}=\Vb^{(k)} \s^{(k-1)}=\begin{bmatrix}
s_1 \\
s_2
\end{bmatrix},\quad \s^{(0)}=\begin{bmatrix}
1 \\
0
\end{bmatrix}.
\end{equation}
Then, in Phase I, Source transmits vector $\sqrt{2P_0}\s^{(k)}$ to the relays over two time-slots, where $P_0$ is the average transmission power per symbol. The channel from Source to the $i$th relay ($\SR_i$) is assumed to be Rayleigh flat-fading and quasi-static during each transmission block. The coefficient of the $\SR_i$ channel is represented by $q_i\sim \CN(0, 1)$. The received signal at the $i$th relay and at the $j$th time-slot is
\begin{equation}
\label{eq:ri}
r_{ij}^{(k)}=\sqrt{2P_0 } \; q_i s_j^{(k)}+ z_{ij}^{(k)},\quad i,j=1,2,
\end{equation}
where $z_{ij}^{(k)} \sim \mathcal{CN}(0,N_0)$ is the noise element at the $i$th relay and $j$th time-slot. The average received SNR per symbol at the relays is $P_0/N_0$. 

Relays 1 and 2 configure their signals as 
\begin{equation}
\label{eq:x1x2}
\begin{split}
x_{1j}^{(k)}&= A\; r_{1j}^{{(k)}}, \quad j=1,2 \\
x_{21}^{(k)}&= -A\;r_{22}^{*^{(k)}}, \quad x_{22}^{(k)}= A\; r_{21}^{*^{(k)}},
\end{split}
\end{equation}
where $A=\sqrt{P_\rt/(P_0+N_0)}$ is the amplification factor and $P_\rt$ is the average transmission power per symbol at the relays. Here, the total power $P$ is allocated between Source and the relays as $P_0=P/2, P_{\rt}=P/4$.

\begin{figure}[t]
\psfrag {km1} [] [] [1.0] {block $(k-1)$}
\psfrag {k} [] [] [1.0] {block $(k)$}
\psfrag {r11} [] [] [.8] {$x_{11}^{(k-1)}$}
\psfrag {r12} [] [] [.8] {$x_{12}^{(k-1)}$}
\psfrag {r13} [] [] [.8] {$x_{11}^{(k)}$}
\psfrag {r14} [] [] [.8] {$x_{12}^{(k)}$}

\psfrag {r21} [] [] [.8] {$x_{21}^{(k-1)}$}
\psfrag {r22} [] [] [.8] {$x_{22}^{(k-1)}$}
\psfrag {r23} [] [] [.8] {$x_{21}^{(k)}$}
\psfrag {r24} [] [] [.8] {$x_{22}^{(k)}$}
\psfrag {T} [t] [] [0.8] {$T_s$}
\psfrag {tau} [] [] [0.8] {$\tau$}
\centerline{\epsfig{figure={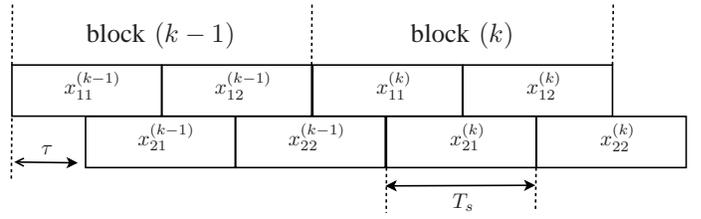},width=9cm}}
\caption{Received signals from Relay 1 and Relay 2 at Destination when Relay 2 is $\tau$ seconds late.}
\label{fig:asyncsig}
\end{figure}

Next, in Phase II, $x_{1j}$ and $x_{2j}$ are simultaneously transmitted from Relays 1 and 2, respectively, to Destination. The channel from the $i$th relay to Destination ($\RiD$) is assumed to be Rayleigh flat-fading and quasi-static during each transmission block. The coefficient of the $\RiD$ channel is represented by $g_i\sim \CN(0, 1)$. The common assumption is that both relays are perfectly synchronized in the symbol level. However, due to the distributed nature of relay networks, relays may have different distances from Destination. Therefore, signals from the relays would arrive at different times. Let us assume that Destination is synchronized with Relay 1 and the signal from Relay 2 arrives $\tau$ seconds later at Destination. This is shown in Fig.~\ref{fig:asyncsig}. It is assumed that $0 \leq \tau \leq T_s,$ where $T_s$ is the symbol duration. 

\begin{figure}[t]

\psfrag {x1} [] [] [0.8] {$x_{11}^{(k)}$}
\psfrag {x2} [] [] [0.8] {$x_{21}^{(k)}$}
\psfrag {x3} [] [] [0.8] {$x_{22}^{(k-1)}$}
\psfrag {Time} [] [] [0.8] {$t/T_s$}
\centerline{\epsfig{figure={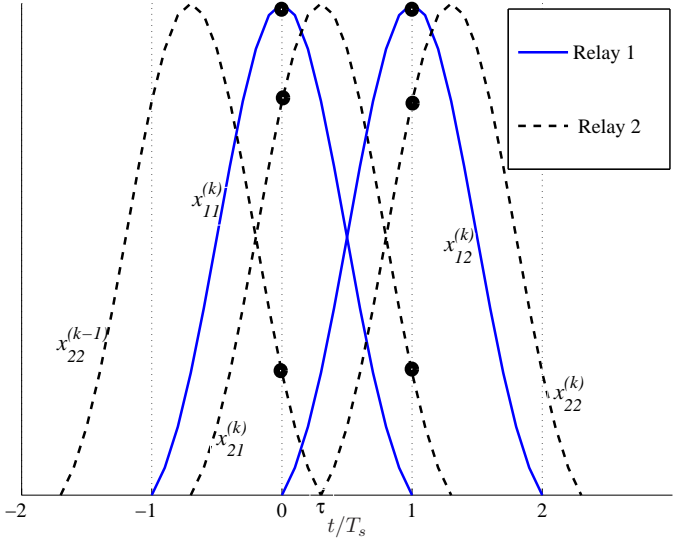},width=9cm}}
\caption{Received signals from Relays 1 and 2 at Destination after the matched-filter using a raised-cosine pulse-shape with roll-off factor $\beta=0.9$ and $\tau=0.3T_s$.}
\label{fig:mf1}
\end{figure}

\addt{The received signal at Destination is passed through a matched-filter and sampled at integer multiples of $T_s$. The baseband signals from the two relays, assuming the raised-cosine pulse-shape and matched-filter $p(t)=\sinc(t/T_s)\cos(\pi \beta t/T_s)/(1-4\beta^2t^2/T_s^2),$ \cite{madhow} with roll-off factor $\beta=0.9$, are depicted in Fig.~\ref{fig:mf1}. As seen in Fig.~\ref{fig:mf1}, the sampled signal after the matched-filter is the super-position of three signals. One signal from Relay 1, whose peak value is at the sampling point, contributes in the sampled signal. Also, depending on $\tau$ and $p(t)$, two fractions of the signal from Relay 2 contribute in the sampled signal.} The received signals at Destination at block-index $(k)$ can be written as
\begin{equation}
\label{eq:yasync}
\begin{split}
y_1^{(k)}=& g_1 x_{11}^{(k)}+ g_{20} x_{21}^{(k)}+ g_{21} x_{22}^{(k-1)}+n_1^{(k)} \\
y_2^{(k)}=& g_1 x_{12}^{(k)}+ g_{20} x_{22}^{(k)}+ g_{21} x_{21}^{(k)}+n_2^{(k)} 
\end{split}
\end{equation}
where $g_{20}=p(\tau) g_{2}, g_{21}=p(T_s-\tau) g_{2}$, and $n_j^{(k)}\sim \CN(0,N_0),\quad j=1,2$ are the noise element at Destination. \addt{Thus, the effect of synchronization error is collected into quantities $g_{20}$ and $g_{21}$. Note that, depending on the number of side-lobes of the pulse-shape filter, more terms may appear in \eqref{eq:yasync}. In our model, the small contributions of the side-lobes of $p(t)$ are neglected.}

By substituting \eqref{eq:ri} and \eqref{eq:x1x2} into \eqref{eq:yasync}, the received signals at Destination in the matrix form can be expressed as
\begin{multline}
\label{eq:y1}
\y^{(k)}= A \sqrt{2P_0} \begin{bmatrix}
s_1 & -s^*_2 \\
s_2 & s^*_1
\end{bmatrix}
\begin{bmatrix}
h_1 \\
h_{20}
\end{bmatrix} \\
+ A\sqrt{2P_0} {h}_{21} \begin{bmatrix}
s_1^{*^{(k-1)}} \\
-s_2^{*^{(k)}}
\end{bmatrix}+\w^{(k)}
\end{multline}
where 
\begin{equation}
\begin{split}
&\y^{(k)}=\left[y_1^{(k)} \quad y_2^{(k)}\right]^t,\\
&h_1=q_1g_1,\; h_{20}=q_2^* g_{20},\; h_{21}=q_2^* g_{21},\\
&\w^{(k)}=\left[w_1^{(k)} \quad w_2^{(k)}\right]^t,\\
&w_1^{(k)}=n_1^{(k)}+A\left(g_1 z_{11}^{(k)}-g_{20}z_{22}^{*^{(k)}}+g_{21}z_{21}^{*^{(k-1)}}\right),\\
& w_2^{(k)}=n_2^{(k)}+A \left(g_1 z_{12}^{(k)}+g_{20} z_{21}^{*^{(k)}}-g_{21}z_{21}^{*^{(k)}}\right).
\end{split}
\end{equation}
As it is seen, in addition to the desired signal, an additional term appears in the system equation. For $\tau=0$, i.e., perfect synchronization, this term is zero and the equivalent noise vector $\w^{(k)}$ is $\CN(\0,\sigma^2\I_2)$, where $\sigma^2=N_0(1+A^2\sum \limits _{i=1}^2|g_i|^2)$ and the average received SNR is $\gamma=A^2 P_0 \sum \limits_{i=1}^{2} |g_i|^2/\sigma^2$, for given $g_1,g_2$.  However, for $\tau>0$, this term would cause a significant ISI and the equivalent noise becomes correlated. If all the channel information and the delay $\tau$ were available at Destination, the transmitted symbols could be jointly decoded. However, if the information is not available (system under consideration), one can treat the ISI as noise. In this case, using \eqref{eq:s[k]} and assuming that channel coefficients are constant during two consecutive blocks, the data symbols can be conventionally decoded as
\begin{equation}
\label{eq:diff-decod}
\hat{v}_1,\hat{v}_2=\arg \min \limits_{\Vc} \|\y^{(k)}-\Vb^{(k)} \y^{(k-1)}\|.
\end{equation}
However, the conventional D-DSTC performs poorly in case of synchronization error as will be shown in Section~\ref{sec:Sim}. 


\section{Proposed Method}
\label{sec:DSTC-OFDM}
In this section, we propose a method for combating the synchronization error in the above system. The method combines differential encoding and decoding with an OFDM approach and is referred to as Differential OFDM (D-OFDM) DSTC. To establish the notation, first a brief review of OFDM systems is provided.

\subsection{OFDM System}
Frequency selective channels are usually modeled with finite impulse response (FIR) filters in the base-band. The channel output is the convolution of the channel impulse response and input sequence which leads to ISI. OFDM is a low complexity approach to deal with the ISI encountered in frequency-selective channels as explained in the following. Let $\{x[n]\}, \quad n=0,\cdots, N-1$ represent the data symbols of length $N$ and $\{h_0,\cdots,h_{L-1}\}$ represent the discrete-time channel of length $L$. The $N$-point IDFT defined as $X[m]=\IDFT\{x[n]\}=1/\sqrt{N}\sum \limits_{n=0}^{N-1}x[n]\exp(j2\pi nm/N),$ is applied to obtain sequence $\{X[m]\},\quad m=0,\cdots,N-1$. Next, a cyclic prefix is appended to the beginning of sequence $\{X[m]\}$ as $\{X[N-L],\cdots,X[N-1],X[0],\cdots,X[N-1]\}$ and the result is injected to the channel. Let us assume that the additive noise is zero. The channel output sequence, after removing the first $L$ received symbols, is $Y[m]=h_l \otimes X[m],\quad m=0,\cdots, N-1$. 
Now, the N-point DFT defined as $y[n]=\DFT\{Y[m]\}=1/\sqrt{N}\sum \limits_{m=0}^{N-1}Y[m]\exp(-j2\pi mn/N)$ is applied to obtain $y[n]=H[n] x[n], \quad n=0,\cdots,N-1$, where $H[n]=\sum \limits_{l=0}^{L-1}h_l \exp(-j2\pi ln/N)$. Using OFDM, the ISI is removed and the $L$-tap frequency-selective channel is converted to $N$ parallel flat-fading channels. 

In the next section, the proposed method is described in detail. 

\subsection{Differential OFDM DSTC}
\begin{figure}[t]
\psfrag {v1} [] [] [0.8] {$v_1[n]$}
\psfrag {v2} [] [] [0.8] {$v_2[n]$}
\psfrag {V} [] [] [0.8] {$\Vb[n]$}
\psfrag {STC} [] [] [0.8] {USTC}
\psfrag {Diff} [] [] [0.8] {Diff.}
\psfrag {Encod} [] [] [0.8] {Encod.}
\psfrag {IDFT} [] [] [0.8] {IDFT}
\psfrag {CP} [] [] [0.8] {Add CP}
\psfrag {s1} [] [] [0.8] {$s_1[n]$}
\psfrag {s2} [] [] [0.8] {$s_2[n]$}
\psfrag {S1} [] [] [0.8] {$S_1[m]$}
\psfrag {S2} [] [] [0.8] {$S_2[m]$}
\centerline{\epsfig{figure={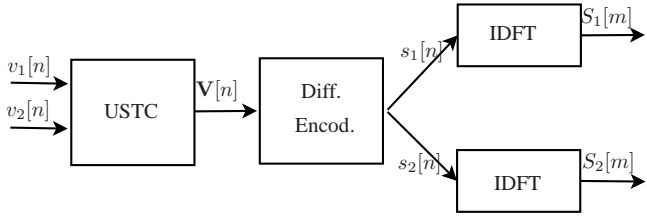},width=8.5cm}}
\caption{Encoding process at Source}
\label{fig:sourceblk}
\end{figure}

Using Eq.~\eqref{eq:yasync}, the effect of synchronization error is modeled by a frequency-selective channel with two taps and the OFDM method is utilized to remove the ISI. Similar to the conventional method, a two-phase transmission process is employed. However, instead of two symbols, a sequence of symbols will be transmitted during each phase.  In Phase I, Source encodes data information as depicted in Fig.~\ref{fig:sourceblk} and transmits $2N$ symbols to the relays. Then, the relays apply a special configuration, append $2L$ symbols and transmit $2(N+L)$ symbols to Destination in Phase II. Finally, Destination removes the $2L$ symbols and decodes the $2N$ symbols.
Transmission of $2N$ symbols from Source to Destination in two phases is referred to as ``one block transmission", indexed by $k\in \Zc$. The description of each step is described in detail as follows.

Let us consider $2N$ data symbols, to be transmitted from Source to Destination, into sequences $\{v_1[n]\},\{v_2[n]\},\quad n=1,\cdots,N$ of length $N$. The two sequences are then encoded to USTC matrices based on \eqref{eq:alamouti} to obtain $\{\Vb[n]\},\quad n=1,\cdots, N$. 
Next, matrices $\{\Vb[n]\}$ are differentialy encoded as 
\begin{equation}
\label{eq:sk-ofdm}
\begin{split}
\s[n]^{(k)}=\Vb[n]^{(k)} \s[n]^{(k-1)}=\begin{bmatrix}
s_{1}[n] \\
s_{2}[n]
\end{bmatrix}, \\
\s[n]^{(0)}=[1 \quad 0]^t, \quad n=0,\cdots,N-1,
\end{split}
\end{equation}
to obtain sequences $\{s_1[n]\}^{(k)},\{s_2[n]\}^{(k)},\quad n=0,\cdots,N-1$ of length $N$. From now on, for simplicity of notations, the block-index $(k)$ is omitted.

Then, the N-point IDFT is applied to $\{s_1[n]\},\{s_2[n]\}$ sequences to obtain $S_1[m]=\IDFT\{s_1[n]\}$ and $S_2[m]=\IDFT\{s_2[n]\},\quad m=0,\cdots,N-1$. The obtained sequences $\{S_1[m]\}$ and $\{S_2[m]\}$ are then transmitted consecutively from Source to the relays over two sub-blocks, in Phase I.

The received signals at the relays for $\quad m=0,\cdots, N-1$ are expressed as
\begin{equation}
\label{eq:Rij}
\begin{split}
R_{ij}[m]=\sqrt{2P_0} q_i S_{j}[m]+Z_{ij}[m],\quad i,j=1,2
\end{split}
\end{equation}
where $Z_{ij}[m] \sim \CN(0,N_0)$ is the noise elements at Relay $i$ and sub-block $j$. 

The received signals at Relays 1 and 2 for $\quad m=0,\cdots,N-1$ are configured as
\begin{equation}
\label{eq:Xij}
\begin{split}
X_{1j}[m]&=A\; R_{1j}[m], \quad j=1,2, \\
X_{21}[m]&=-A \; \accentset{\circ}{R}_{22}^*[m], \\
X_{22}[m]&=A \;  \accentset{\circ}{R}_{21}^*[m],
\end{split}
\end{equation}
where $A$ is the amplification factor and $\accentset{\circ}{R}_{2j}[m]$ is the \emph{circular time-reversal} \cite{DSP-Boaz} of $R_{2j}[m]$ defined as
\begin{equation}
\label{eq:CTR}
\accentset{\circ}{R}_{2j}[m]= \left\lbrace 
\begin{matrix}
R_{2j}[0], & m=0 \\
R_{2j}[N-m], & \mbox{otherwise}.
\end{matrix} \right.
\end{equation} 
Before transmission, the last $L$ symbols of sequences $\{X_{ij}[m]\}, i,j=1,2$ are appended to their beginnings as the cyclic prefix to obtain $$\{X_{ij}[N-L],\cdots,X_{ij}[N-1],X_{ij}[0],\cdots,X_{ij}[N-1]\}$$ with length $N+L$. Here, $L$ is the cyclic prefix length determined based on the amount of delay between the received signals from both relays and will be discussed shortly. 

In Phase II, Relay 1 transmits sequences $\{X_{11}[m]\}$ and $\{X_{12}[m]\}$, while Relay 2 transmits $\{X_{21}[m]\}$ and $\{X_{22}[m]\}$, for $m=-L,\cdots,0,\cdots,N-1$, during two consecutive sub-blocks or $2(N+L)$ symbols, to Destination. 
 
At Destination, without loss of generality, let us assume that the received signal from Relay 2 is $(d T_s+\tau)$ seconds delayed with respect to that of Relay 1, where $d$ is an integer number and $0 \leq \tau\leq T_s$. Thus, to avoid ISI, the cyclic-prefix length is determined as $L>d$. If the delay, as shown in Fig.~\ref{fig:asyncsig}, is less than one symbol duration, $L=1$ is enough. In practice the relays do not need to know the delay and, based on the propagation environment, the maximum value of $d$ in the network can be estimated and used to determine the cyclic prefix length.

In this case, the received signals during two sub-blocks, after removing the first $L$ symbols, can be expressed as
\begin{equation}
\begin{split}
Y_1[m]&=g_1 X_{11}[m]+g_{20} X_{21}[m-d]
\\&+ g_{21} X_{21}[m-1-d]+ W_{1}[m] 
\\&= g_1 X_{11}[m]+\left(g_{2}[m] \otimes X_{21}[m-d]\right)+W_1[m],
\\& \quad m=0,\cdots,N-1
\end{split}
\end{equation}

\begin{equation}
\begin{split}
Y_2[m]&=g_1 X_{12}[m]+g_{20} X_{22}[m-d]
\\&+g_{21} X_{22}[m-1-d]+ W_{2}[m] 
\\&= g_1 X_{12}[m]+\left(g_{2}[m]  \otimes X_{22}[m-d]\right)+W_2[m],
\\& \quad m=0,\cdots,N-1
\end{split}
\end{equation}
where $g_{2}[m]=\sum \limits_{l=0}^1 g_{2l} \delta[m-l]$.

By substituting \eqref{eq:Xij} and \eqref{eq:Rij} into the above equations, one obtains, for $m=0,\cdots,N-1$
\begin{multline}
Y_1[m]=A\sqrt{2P_0} \left(h_1 S_1[m]-h_{2}[m] \otimes \accentset{\circ}{S}_2^*[m-d]\right)\\+A \left( g_1 Z_{11}[m]-  g_{2}[m] \otimes \accentset{\circ}{Z}_{22}^*[m-d]\right)+W_1[m],
\end{multline}

\begin{multline}
Y_2[m]=A\sqrt{2P_0} \left(h_1 S_2[m]+h_{2}[m] \otimes \accentset{\circ}{S}_1^*[m-d]\right)\\+A \left(g_1 Z_{12}[m]+g_{2}[m] \otimes \accentset{\circ}{Z}_{21}^*[m-d]\right)+W_2[m],
\end{multline}
where sequences $\accentset{\circ}{S}_j[m]$ and $\accentset{\circ}{Z}_{2j}[m]$ are the circular time-reversal of sequences $S_j[m]$ and $Z_{2j}[m]$, respectively, as defined in \eqref{eq:CTR}. $h_1=q_1 g_1, h_2[m]=\sum \limits_{l=0}^1 h_{2l}\delta[m-l], h_{20}=q_2^*g_{20}, h_{21}=q_2^*g_{21}$, and as defined in Section~\ref{sec:cdd}.  

By taking the N-point DFT of $Y_1[m],Y_2[m]$ sequences and the properties of circular time-reversal\footnote{If $S[m]=\IDFT\{s[n]\}$ then $S^*[m]=\DFT\{s^*[n]\}$. Also, circular time-reversal operation converts DFT to IDFT and vice versa, i.e., if $\accentset{\circ}{S}^*[m]$ is the circular time-reversal of sequence $S^*[m]=\DFT\{s^*[n]\}$ then $\accentset{\circ}{S}^*[m]=\IDFT\{s^*[n]\}$ \cite{DSP-Boaz}.} sequences, for $n=0,\cdots,N-1,$  one derives 
\begin{equation}
\begin{split}
y_1[n]&= A\sqrt{2P_0} \left(h_1 s_{1}[n]-H_2[n] s_{2}^*[n]\right)+\widetilde{w}_1[n],\\
y_2[n]&= A\sqrt{2P_0} \left(h_1 s_{2}[n]+H_2[n] s_{1}^*[n]\right)+\widetilde{w}_2[n],\\
\end{split}
\end{equation}
with
\begin{equation}
\begin{split}
H_2[n]&=q_2^* G_2[n],\\
G_2[n]&=\left(g_{20}+g_{21}\et^{-j2\pi n/N}\right)\et^{-j2\pi d/N},\\
\widetilde{w}_1[n]&=A \left( g_1 z_{11}[n]- G_{2}[n] z_{22}^*[n]  \right)+w_1[n],\\
\widetilde{w}_2[n]&=A \left( g_1 z_{12}[n]+ G_{2}[n] z_{21}^*[n]  \right)+w_2[n],\\
z_{11}[n]&=\DFT\{Z_{11}[m]\},\; z_{22}[n]=\DFT\{{Z}_{22}[m]\},\\
z_{12}[n]&=\DFT\{Z_{12}[m]\},\;z_{21}[n]=\DFT\{{Z}_{21}[m]\},\\
w_1[n]&=\DFT\{W_{1}[m]\},\; w_2[n]=\DFT\{W_{2}[m]\}.
\end{split}
\end{equation}
Clearly, $z_{ij}[n]\sim \CN(0,N_0)$ and $w_{j}[n]\sim \CN(0,N_0)$ for $i,j=1,2$.

The received signals for the block-index $(k)$, $\y[n]^{(k)}=[y_1[n]\quad y_2[n]]^t$, in the matrix form, for $0\leq n\leq N-1$, can be expressed as
\begin{equation}
\label{eq:y-ofdm}
\y[n]^{(k)}=A \sqrt{2P_0} \begin{bmatrix}
s_{1}[n] & -s_{2}^*[n] \\
s_{2}[n] & s_{1}^*[n]
\end{bmatrix}
\begin{bmatrix}
h_1 \\
H_2[n]
\end{bmatrix}+
\begin{bmatrix}
\widetilde{w}_1[n] \\
\widetilde{w}_2[n]
\end{bmatrix}.
\end{equation}
It is pointed out that, for given $g_1,g_2$, the equivalent noise $[\widetilde{w}_1[n]\quad \widetilde{w}_2[n]]^t\sim \CN(\0,\sigma^2[n]\I_2)$, where
\begin{align}
\label{eq:cn}
\sigma^2[n]&=N_0(1+A^2(|g_1|^2+|g_2|^2c[n]),\\
c[n]&= \lvert p(\tau)+p(T_s-\tau) \et^{-j2\pi n/N} \rvert^2.
\end{align}
Also, the received SNR per symbol, for given $g_1,g_2$ can be obtained as 
\begin{equation}
\label{eq:gama}
\gamma[n,\tau]= \frac{A^2P_0 (|g_1|^2+|g_2|^2c[n])}{N_0\left(1+A^2(|g_1|^2+|g_2|^2c[n])\right)}.
\end{equation}

\begin{figure}[t]
\psfrag {n} [] [] [1] {$n$}
\psfrag {gama} [] [] [1] {$\gamma[n,\tau]$, dB}
\centerline{\epsfig{figure={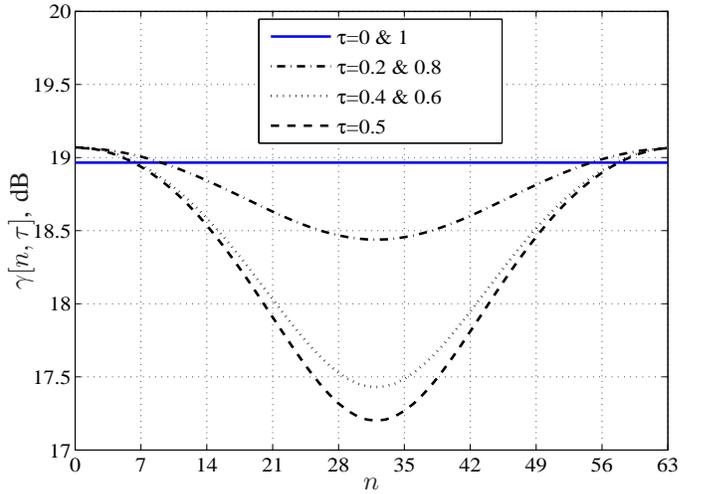},height=6.5cm,width=9cm}}
\caption{Average received SNR vs. $n$ and $\tau$, $N=64,L=1$, $P/N_0=25$dB, $P_0=P/2,P_\rt=P/4$, $|g_1|^2=|g_2|^2=1$.}
\label{fig:fig:gama}
\end{figure}

With the raised-cosine filter defined in Section~\ref{sec:cdd}, $p(\tau)=1$ and $p(T_s-\tau)=0$ for $\tau=0$ and hence $c[n]=1$. Thus, the noise variance and the received SNR of the proposed system are the same as that of the conventional D-DSTC for $\tau=0$. However, for $\tau \neq 0$ the average received SNR is a function of $\tau$ and $n$. To see this dependency, $\gamma[\tau,n]$ is plotted versus $n$ and $\tau$ in Fig.~\ref{fig:fig:gama}, when $N=64,L=1$, $P/N_0=25$dB, $P_0=P/2,P_\rt=P/4$ and for simplicity $|g_1|^2=|g_2|^2=1$. As can be seen, $\gamma[\tau,n]$ is symmetric around its minimum at $n=N/2-1$. Also, overall, $\gamma[\tau,n]$ decreases with increasing $\tau$ and reaches its minimum value at $\tau=0.5T_s$. Then it increases with increasing $\tau$ towards $T_s$ such that $\gamma[\tau,n]=\gamma[T_s-\tau,n]$. This phenomena yields the same average BER for symmetric values of $\tau$ around $0.5 T_S$, as will be seen in the simulation results.

By writing \eqref{eq:y-ofdm} for two consecutive block-indexes $(k),(k-1)$, using \eqref{eq:sk-ofdm} and assuming that $h_1$ and $H_2[n]$ are constant over two consecutive blocks, the differential decoding is applied for $\quad n=0,\cdots,N-1$
\begin{equation}
\label{eq:decod-ofdm}
\hat{v}_1[n],\hat{v}_2[n]=\arg \min \limits_{\Vc} \|\y[n]^{(k)}-\Vb[n]^{(k)} \y[n]^{(k-1)}\|,
\end{equation}
to decode the $2N$ data symbols. Because of the orthogonality of $\Vb[n]$, symbols $v_1[n],v_2[n]$ are decoded independently, without any knowledge of CSI or delay. It is easy to see that, due to the structure of Eq.\eqref{eq:y-ofdm}, the desired diversity of two is achieved in this system.

\section{Simulation Results}
\label{sec:Sim}
In this section the relay network in Fig.~\ref{fig:sysmodel} is simulated in various scenarios. Through these simulations, the effectiveness of the proposed method against synchronization error is illustrated and compared with conventional D-DSTC \cite{D-DSTC-Y} and coherent DSTC \cite{DSTC-HJ}. 

The channel coefficients $q_1,q_2,g_1,g_2$, are assumed to be static during each OFDM block and change from block to block according to the Jakes' model with the normalized Doppler frequency of $f_DT_s=10^{-3}$. The simulation method of \cite{ch-sim} is used to generate the channel coefficients. BPSK modulation is used to convert information bits into symbols. Also, $N=64$-point DFT and IDFT with a cyclic prefix length of $L=1$ are employed in the simulation. The system is simulated for various amounts of delay $\tau=(0, 0.2,0.4,0.6,0.8,1)T_s$ and $T_s=1$. 

\begin{figure}[t]
\psfrag {tau} [] [] [0.8] {$\tau$}
\psfrag {BER} [] [] [0.8] {BER}
\psfrag {Total Power} [] [] [0.8] {$P/N_0$dB}
\centerline{\epsfig{figure={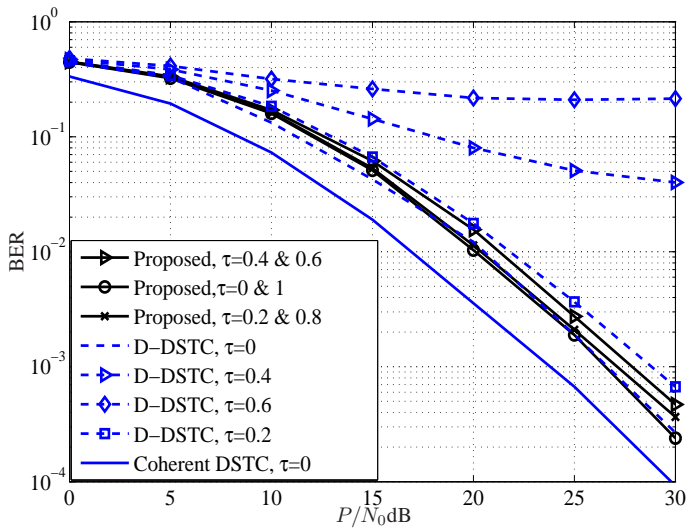},height=7cm,width=9cm}}
\caption{Simulation for BER of D-OFDM DSTC (proposed method, $N=64, L=1$), D-DSTC \cite{D-DSTC-Y}, and coherent DSTC\cite{DSTC-HJ} using BPSK for various values of delays.}
\label{fig:ofdm_m2}
\end{figure}

Fig.~\ref{fig:ofdm_m2} depicts the BER results of the D-OFDM DSTC system versus $P/N_0$, where $P$ is the total power in the network. For comparison purposes, the BER results of the conventional D-DSTC system \cite{D-DSTC-Y} are also added to the figure for various values of $\tau$. Moreover, the BER curve of coherent DSTC \cite{DSTC-HJ} with perfect synchronization $\tau=0$ is plotted as a benchmark. 

As shown in the figure, the performance of the conventional D-DSTC system is severely degraded for $\tau>0.2$ and an error floor appears in the BER curves. On the other hand, the proposed method is able to deliver the desired performance for all values of the delays. As explained in Section~\ref{sec:DSTC-OFDM}, the BER curves are symmetric around $\tau=0.5T_s$. Hence the BER curves of $\tau=0.2T_s$ and $\tau=0.8T_s$ and of $\tau=0.4T_s$ and $\tau=0.6T_s$ are the same. Moreover, the BER curves of $\tau=0, T_s$ of the proposed method are similar to that of the conventional D-DSTC with $\tau=0$, as expected. A performance difference of around 3 dB is seen between coherent DSTC with perfect synchronization and that of D-OFDM DSTC.

\section{Conclusion}
\label{sec:con}
While collecting channel information is challenging, synchronization error is also inevitable in distributed space-time relay networks. Hence, in this paper a method was proposed that does not require any channel information and is very robust against synchronization error. The method combines differential encoding and decoding with an OFDM-based approach to circumvent channel estimation and deal with synchronization error. It was shown through simulations that the method works well for various synchronization error values.

\balance
\bibliographystyle{IEEEbib}
\bibliography{ref/references}

\end{document}